\documentclass[aps,pre,twocolumn,groupedaddress,showpacs]{revtex4}
\usepackage{amssymb}
\usepackage{graphicx,color}
\definecolor{r}{rgb}{1,0,0}
\definecolor{g}{rgb}{0,1,0}
\definecolor{b}{rgb}{0,0,1}

\begin{document}


\title{Centrifugal compression of soft particle packings -- theory and experiment}


\author{K. N. Nordstrom$^1$, E. Verneuil$^{1,2}$, W. G. Ellenbroek$^1$, T. C. Lubensky$^1$, J. P. Gollub$^{1,3}$ and D. J. Durian$^1$}
\affiliation{$^{1}$Department of Physics and Astronomy, University of Pennsylvania, Philadelphia, PA 19104-6396, USA}
\affiliation{$^{2}$Complex Assemblies of Soft Matter, CNRS-Rhodia-UPenn UMI 3254, Bristol, PA 19007-3624, USA}
\affiliation{$^{3}$Department of Physics and Astronomy, Haverford College, Haverford, PA 19041-1392, USA}


\date{\today}

\begin{abstract}
An exact method is developed for computing the height of an elastic medium subjected to centrifugal compression, for arbitrary constitutive relation between stress and strain.  Example solutions are obtained for power-law media and for cases where the stress diverges at a critical strain -- for example as required by packings composed of deformable but incompressible particles.  Experimental data are presented for the centrifugal compression of thermo-responsive N-isopropylacrylamide (NIPA) microgel beads in water.  For small radial acceleration, the results are consistent with Hertzian elasticity, and are analyzed in terms of the Young elastic modulus of the bead material.  For large radial acceleration, the sample compression asymptotes to a value corresponding to a space-filling particle volume fraction of unity.  Therefore we conclude that the gel beads are incompressible, and deform without deswelling.  In addition, we find that the Young elastic modulus of the particulate gel material scales with cross-link density raised to the power $3.3\pm0.8$, somewhat larger than the Flory expectation.
\end{abstract}

%


\maketitle



\def\es{\sigma_{zz}}   


Interest in colloidal suspensions has been spurred by the advent of optical tools to image and manipulate behavior at the particle scale \cite{VanblaaderenScience95, Crocker96, GrierNature03, WeeksJPCM07}.  While much research focusses on hard-sphere and charged systems, other work concerns the behavior of elastic microgel particles composed of a swollen polymer network such as N-isopropylacrylamide (NIPA) \cite{RichteringJCP99, SaundersACIS99, PeltonACIS00, HeyesSM09}.  Since such particles are soft and can swell/de-swell in response to variation of temperature, pH, or salt concentration, they are ideal as model systems for experiments at very high volume fractions where the particles are pressed together so that the material behaves as a jammed solid-like paste \cite{CloitrePRL00, CloitrePRL03, CloitreJR06, TrappePT09, WeitzNature09, YodhNature09, YodhPRL10, Nordstrom2010}.  

The mechanical behavior of a jammed packing ultimately originates in the elastic nature of the constituent particles.  For microgel particles this depends on crosslink density and swelling state, and ought to form a crucial part of sample characterization.  Similar considerations apply to packing of bubbles, cells, grains, etc.  For large enough particles, the deformation of individual particles may be visualized under applied load -- either of a packing \cite{LachhabEPJB1999, BehringerNature05, BandiPreprint2009} or of a gel single bead \cite{AndreiJDCP1996, KnaebelPGN1997, KnaebelPGN1997b, EgholmJAPS2006}.   However, this is not feasible for submicron-scale colloidal microgel beads.  

In this paper, we demonstrate how the elasticity of a medium may be characterized by centrifugal compression, and we illustrate our method with experiments on $\approx 1~\mu$m diameter NIPA microgel beads.   Not surprisingly, theories for centrifugal compression have been proposed earlier.  Ref.~\cite{BuscallColloids1982} approximates the overall sample compression in terms of an average pressure across the medium.  Ref.~\cite{AksayJACS94} computes an approximate compression profile for power-law media.  By contrast we develop an {\it exact} prediction for sample height versus radial acceleration, for arbitrary stress versus strain constitutive relation.  Our general theory applies to packings of small as well as large particles and, also, to any elastic medium such as an aggregated suspension \cite{ZukoskiJACS96, WeitzEPJE09} that can be compressed either by centrifugation or gravity.  Nevertheless our primary interest here is in a non-cohesive random packing of spheres, for which we discuss how to relate bulk to particle-scale elasticity in light of recent advances regarding non-affine deformation.

\section{Theory of sample compression}

\begin{figure}
\includegraphics[width=2.5in]{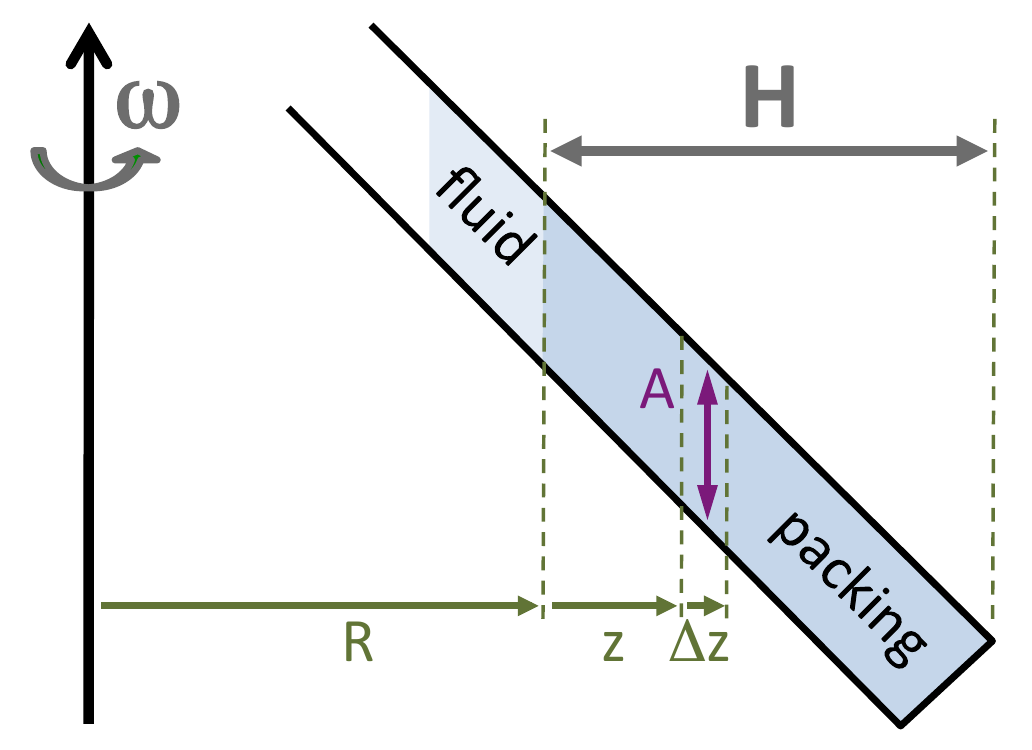}
\caption{(Color online)
Schematic depiction of a soft particle packing under centrifugal compression.  For rotation at angular speed $\omega$ around the vertical axis as labeled, it comes to mechanical equilibrium at radial ``height'' $H$ above the bottom of the sample.  For this particular sample and centrifuge geometry, the cross sectional area $A$ at depth $z$ below the supernatant fluid-packing interface is constant except at the very bottom.}
\label{setup}
\end{figure}

Three main ingredients are required in order to predict sample height $H$ versus angular rotation speed $\omega$ for analysis of experimental data.  This includes statements of mass conservation and force balance, as well as a constitutive model for compressive stress versus strain based on the elastic nature of the particle packing.  Some of the key quantities for this task are defined on the schematic diagram of the experiment in Fig.~\ref{setup}.  The sample itself consists of $N$ particles in a suspending fluid of density $\rho_f$.  To encompass all possibilities of particle deformation and deswelling, we define $m$ and $v$ respectively as the mass and volume of particulate material in each particle.  For example if the particles are incompressible homogeneous droplets, then $m$ is the mass of the droplet and $v$ is its volume.  But if the particles are porous, then $m$ and $v$ do not include contributions from the suspending fluid within the pores.  Therefore, in any case, $m$ and $v$ and the particulate material mass density $\rho_p=m/v$ are all constant and do not change no matter how the particles deform or deswell.

In the limit of vanishing $\omega$, the particles are unstrained but close-packed at a volume fraction $\phi_c$ and a number density $n_c$ that are constant throughout the volume of the packing. The packing extends a radial distance or ``height'' $H_c$ inwards from the bottom of the container, with the supernatant fluid ``above'' -- closer to the rotation axis.  At nonzero $\omega$, the particle packing compresses to a smaller height $H$, and the strain $\gamma$ varies with the radial ``depth'' $z$ below the supernatant fluid, so that everywhere the elastic stress gradient counteracts the centrifugal buoyancy pressure gradient.  In many centrifuges the sample tube is not perpendicular to the rotation axis, as in the depiction of Fig.~\ref{setup}.  It is important to emphasize that the distance $R$ between the rotation axis and the supernatant-packing interface, the ``height'' $H$, and the  ``depth'' $z$, are all measured radially and not along the length of the tube.

\subsection{Mass conservation}

A thin slice of unstrained sample, with area $A$ and thickness $L$, contains $n_c A L$ beads.  If compressed to $L-\Delta L$, the number density $n$ increases but the number of beads $n A (L-\Delta L)$ is unchanged.  Therefore the number density is
\begin{equation}
 n = {n_c  \over 1 - \gamma}
\label{MassConsLocal}
\end{equation}
where $\gamma = \Delta L / L$ is the compressive strain.  If the entire sample has constant area $A$, and is compressed in radial height from $H_c$ to $H$ under centrifugation, then the total number of beads in the entire sample is similarly unchanged, $n_c A H_c = \int_0^H n A {\rm d}z$.  The global expression of mass conservation is thus
\begin{equation}
    H_c = \int_0^H {{\rm d}z \over 1-\gamma}.
\label{MassConsGlobal}
\end{equation}
More generally if the cross-sectional area varies across the sample, for example as at the bottom of the tube depicted in Fig.~\ref{setup}, then the expression of mass conservation is $V_c = \int_0^H [A/(1-\gamma)]{\rm d}z$.  In our experiments, we use sample tubes of very large length to diameter ratio, so that $A$ can be considered as uniform and Eq.~(\ref{MassConsGlobal}) is accurate.   Note that all these expressions of mass conservation hold whether sample compression is due to deformation of particles at constant volume, or deswelling of particles, or some combination of both.

\subsection{Force balance}

The particles in the thin slice of volume $A\Delta z$ depicted in Fig.~\ref{setup} rotate with radius $R+z$ and hence must experience a net force that points radially inward and that equals total particle mass $mnA\Delta z$ times radial acceleration $a_r=\omega^2 (R+z)$.  Assuming that the sample boundary is frictionless, this net force arises both from the elasticity of the packing and from centrifugal buoyancy.  The compressive strain $\gamma$ and elastic stress $\es$ increase with radial ``depth" $z$, so the elastic force on the thin slice is $A ({\rm d}\es /{\rm d}z)\Delta z$.  The mass of fluid displaced by particulate material within the slice is $\rho_f nv A \Delta z$, so the centrifugal buoyant force is given by Archimedes as this times the radial acceleration.  Altogether, the statement of Newton's second law is
\begin{equation}
  (\rho_f nv A \Delta z)a_r + A{ {\rm d}\es \over {\rm d}z }\Delta z = (nmA\Delta z)a_r.
\label{newton}
\end{equation}
Here the right-hand side is mass of particles in the slice $A\Delta z$ times their acceleration, and the left-hand side is the sum of forces acting on the particles; the first force is due to the surrounding fluid and the second is due to the surrounding particles.  Note that the factor $A\Delta z$ cancels, so that the only depth-dependent terms are the number density, given by Eq.~(\ref{MassConsLocal}) as $n=n_c/(1-\gamma)$, the elastic stress gradient ${\rm d}\es /{\rm d}z$, and the radius of motion $(R+z)$ appearing in $a_r$.  Also, the constant $n_c (m-\rho_f v)$ is identified as $\phi_c\Delta\rho$, where $\phi_c$ is the volume fraction occupied by the particles at close-packing and $\Delta \rho$ is the density difference between particle and fluid material.  Thus the final expression for force balance simplifies to
\begin{equation}
  { {\rm d}\es \over {\rm d}z } = {  \phi_c \Delta\rho \omega^2(R+z) \over 1-\gamma }.
\label{fullforcebalance}
\end{equation}
for any container shape and for any combination of particle deformation and/or deswelling.

\subsection{Formal solution}

As a constitutive model we suppose that the elastic stress of the compressed particle packing may be written in general as
\begin{equation}
    \es = Y s(\gamma)
\label{GenPress}
\end{equation}
where $Y$ is a materials property with dimensions of force per area and $s(\gamma)$ is a dimensionless function of the compressive strain $\gamma$.  We also suppose that sample height is small compared to the centrifuge radius, $y \le H\ll R$.  Then the force balance equation (\ref{fullforcebalance}) becomes
\begin{eqnarray}
    {{\rm d}s \over {\rm d}z} &\approx& {k \over 1-\gamma}, \label{fb}\\
    k &\equiv& \phi_c \Delta\rho \omega^2 R/Y. \label{k}
\end{eqnarray}
Note that $k$ is defined as a reciprocal length that characterizes both the materials and the radial acceleration.  With a change of variables, the force balance equation may be integrated by parts:
\begin{eqnarray}
 \int_0^z k{\rm d}z &=&   \int_0^{\gamma(z)} (1-\gamma){{\rm d}s\over {\rm d}\gamma}{\rm d}\gamma, \label{parts1}\\
   kz   &=&   [1-\gamma(z)]s+\int_0^{\gamma(z)} s{\rm d}\gamma, \label{parts2}\\
    &\equiv& kf(s) \label{parts3}.
\end{eqnarray}
For a specific form of $s(\gamma)$ characterizing the elastic nature of the packing, Eq.~(\ref{GenPress}), the integration in Eq.~(\ref{parts2}) is to be performed and the result is to be expressed not in terms of $\gamma$ but rather in terms of $s$; this defines the function $f(s)$ in Eq.~(\ref{parts3}) and its inverse $f^{-1}(z)$.

Finding the function $f(s)$ defined by Eqs.~(\ref{parts2}-\ref{parts3}) is tantamount to finding the height of the sample, if the cross-sectional area is constant.  This can be seen by using the force balance equation~(\ref{fb}) to re-express the mass conservation equation~(\ref{MassConsGlobal}):
\begin{eqnarray}
    H_c &=& \int_0^H {{\rm d}z \over 1-\gamma}, \label{soln1}\\
            &=& \int_0^{f^{-1}(H)} { {\rm d}s \over k }, \label{soln2}\\
            &=& {1\over k}f^{-1}(H). \label{soln3}
\end{eqnarray}
Multiplying by $k$ and taking the inverse gives the final formal result for radial packing height as a function of rotation speed:
\begin{equation}
    H = f(kH_c).
\label{soln}
\end{equation}
To recap, this solution assumes that the sample container has a constant cross-sectional area, and that the radius of circular motion is large compared to the packing height.  However it makes no assumptions about whether sample compression is due to deformation or deswelling of the particles.  And it does not rely on explicit computation of strain versus radial depth.

\subsection{Power-law elasticity}

In this and the following two sub-sections, we use the above formalism to predict sample height versus radial acceleration for three specific stress-strain constitutive relations of potential experimental interest.  The first and simplest is the elastic stress is a power-law of the strain,
\begin{equation}
    \es = Y\gamma^a,
\label{Ppower}
\end{equation}
so that the dimensionless function defined by Eq.~(\ref{GenPress}) is $s(\gamma)=\gamma^a$.  The value $a=1$ corresponds to a linear spring-like medium and the value $a=3/2$ corresponds to a packing of Hertzian spheres with small deformations.  For the general case it is straightforward to carry out the integration in Eq.~(\ref{parts2}) and simplify to $kz=\gamma^a[1-\gamma a/(1+a)]$.  It is also straightforward to invert for $\gamma=s^{1/a}$ and re-express as $kz=s[1-s^{1/a}a/(1+a)]$, the right-hand side of which defines $kf(s)$.  According to Eq.~(\ref{soln}) the sample height is thus $H=f(H_ck)=(1/k)(H_ck)[1-(H_ck)^{1/a}a/(1+a)$, which we express as
\begin{equation}
    {H \over H_c} = 1 - {a \over 1+a}\left(H_ck\right)^{1/a}.
\label{Hpower}
\end{equation}
Since $k\propto\omega^2R$, by definition in Eq.~(\ref{k}), the fractional decrease in sample height, and also the average strain of the entire sample $\langle \gamma \rangle = 1-H/H_c$, vary with the experimentally-accessible control parameters as a linear function of $(\omega^2RH_c)^{1/a}$ and a proportionality constant that depends on materials parameters.  Though Eq.~(\ref{Hpower}) is remarkably simple, we emphasize that it is an {\it exact} solution, good for any radial acceleration or amount of sample compression.

The form of Eqs.~(\ref{soln},\ref{Hpower}), and also of the predictions of sample height in the following sections for different constitutive laws, suggest that data may be conveniently analyzed in terms of a plot of $H/H_c$ versus the length $x=\omega^2 R H_c/g$.  This should cause data collapse for different initial sample heights, and thus serves as a nice check.  Then fits may be made to $H/H_c=1-[a/(1+a)] (\kappa x)^{1/a}$ where the fitting parameter $\kappa$ is an inverse length defined so that $H_ck = \kappa x$.   According to this definition and Eq.~(\ref{k}), the elastic constant of the bulk medium is then
\begin{equation}
	Y=\phi_c \Delta\rho g/\kappa
\label{Ykappa}
\end{equation}
The quality of the fit is bound to be good for small compression, and to give a reliable value for $Y$.  Since Eq.~(\ref{Hpower}) is an exact solution, any deviation of the fit from the data at larger compression is directly related to a deviation of the actual constitutive relation from $\es = Y\gamma^a$.  This makes centrifugal compression a sensitive probe of the elastic nature of the packing.  For example, the sample height for an actual sample will surely not decrease to zero as predicted by  Eq.~(\ref{Hpower}) for strong but finite centrifugation, due to stiffening at large strains.

\subsection{Linear with maximum strain}

There must exist a maximum strain $\gamma_m$ at which all fluid is expelled and the packing is pure particulate material.  If the particulate material is incompressible, then the elastic stress must diverge at this maximum strain.  A simple constitutive law that is linear at small strain and that diverges at $\gamma_m$ is
\begin{equation}
    \es = {Y \gamma \over 1 - \gamma/\gamma_m}.
\label{Plinearmax}
\end{equation}
Though {\it ad-hoc}, this form has the dual virtue of being invertible for strain versus stress and of being integrable in Eq.~(\ref{parts2}).  Taking advantage of these features, the resulting packing height is computed exactly to be
\begin{equation}
    {H\over H_c} = 1 - \gamma_m + {{\gamma_m}^2\over H_ck}\ln \left[1 + { H_c k \over \gamma_m} \right],
\label{Hlinearmax}
\end{equation}
where the reciprocal length $k$ is defined by Eq.~(\ref{k}), as before.  For gentle centrifugation, $H_c k/\gamma_m \ll 1$, this may be expanded as $H/H_c = 1-(H_ck)/2 + (H_ck)^2/(3\gamma_m)$ plus higher order terms; note that the leading behavior is identical to Eq.~(\ref{Hpower}) for $a=1$.  For strong centrifugation, $H_c k/\gamma_m \gg 1$, the average strain of the entire sample approaches $\gamma_m$ as $H/H_c = 1 - \gamma_m + \gamma_m\ln[H_ck/\gamma_m]/(H_ck/{\gamma_m})+\gamma_m/(H_ck/\gamma_m)^2$ plus higher order terms.  This prediction may be a reasonable expectation for emulsions, for which the droplets are incompressible and are often assumed to interact as repulsive linear springs.

\subsection{Hertzian with maximum strain}

For a packing of solid spherical particles that are elastic but incompressible, we reason as above that the stress must be Hertzian at low strains and must diverge at some finite maximum strain $\gamma_m$.  The simplest such form we can conceive, that is also analytically tractable within the context of the formal solution for packing height vs compression, is 
\begin{equation}
    \es = {Y \gamma^{3/2} \over 1 - (\gamma/\gamma_m)^{3/2}}.
\label{Phertzianmax}
\end{equation}
As above, this expression may be inverted for strain versus stress, and it also may be integrated in Eq.~(\ref{parts2}).  The resulting packing height is computed exactly to be
\begin{eqnarray}
    {H\over H_c} &=& 1 - {\gamma_m \over W} + {{\gamma_m}^{5/2} \over 3 H_c k} \bigg( \pi/\sqrt{3} \cr
               & &  -2\sqrt{3}\tan^{-1}\left[ (1 + 2 W)/ \sqrt{3})\right] \cr
               & &  +\ln\left[(1+W+W^2)/(1-W)^2\right] \bigg), \label{Hhertzianmax}\\
    W &=& \left( {H_c k \over {\gamma_m}^{3/2} + H_c k }\right)^{1/3}. \label{W}
\end{eqnarray}
The reciprocal length $k$ is defined by Eq.~(\ref{k}), as before.  For gentle centrifugation, $H_c k/\gamma_m \ll 1$, the sample height prediction behaves as $H/H_c = 1-(3/5)(H_ck)^{2/3} + (H_ck)^{5/3}/(4{\gamma_m}^{3/2})$ plus higher order terms; note that the leading behavior is identical to Eq.~(\ref{Hpower}) for $a=3/2$.  For strong centrifugation, it may also be verified that the sample compression asymptotes to $H/H_c=1-\gamma_c$, as expected.

\subsection{Sphere and packing elasticities}

In this last subsection on theory, we discuss the connection of microscopic particle properties to the macroscopic stress-strain relations for the case of elastic spheres.   In particular, how does the measured value of the parameter $Y$ defined by $\es = Y \gamma^{3/2}$ depend upon the elasticity of the sphere material?  To begin we recall the classic calculation by Hertz for two elastic spheres of equal diameter $d$, brought into contact such that their centers are a distance $h$ closer together than the sum of their radii.  For small deformation, the repulsive force is
\begin{equation}
    F = {E d^2 \over 3(1-\nu^2)}\left({h\over d}\right)^{3/2},
\label{Fhertz}
\end{equation}
where $E$ is the Young elastic modulus and $\nu$ is the Poisson ratio of the sphere material.  A derivation of this expression is given by Landau and Lifshitz \cite{LandauElasticity}, where it culminates in their Eq.~(9.15).  It is also given by Walton \cite{Walton87}, where it culminates in his Eqs.~(2.5~\&~2.15), expressed as combinations of Lam\'e parameters that reduce to $B=(1-\nu^2)/(\pi E)$ and $C=\nu(1+\nu)/(\pi E)$.

The main contribution in Walton's paper \cite{Walton87} is computation of the effective elastic moduli of a random packing of non-cohesive elastic spheres, assuming affine deformation and averaging over a fixed set of randomly-oriented contacts at which the actual stress and strain fields are computed.  The pressure required to achieve uniform compression $\gamma_{ii}=\gamma$ is given in his Eq.~(3.19) as
\begin{eqnarray}
    P &=& {\phi Z E \over 3\pi(1-\nu^2)} \gamma^{3/2}, \label{Pwalton1}\\
        &\approx& {Z E \over 9\pi\sqrt{3\phi_c}(1-\nu^2)} (\phi-\phi_c)^{3/2}, \label{Pwalton2}
\end{eqnarray}
where $\phi$ is the volume fraction of spheres and $Z$ is the average number of contacts per sphere.  For ``perfectly smooth [{\it sic}]'' frictionless spheres that support no shear traction across the contact area, the stress for uniaxial compression $\gamma_{zz}=\gamma$ is given in Walton's Eq.~(3.26) as
\begin{eqnarray}
    \es &=& {\phi Z E\over 6 \pi (1-\nu^2)} \gamma^{3/2} \equiv Y\gamma^{3/2}, \label{Ywalton1}\\
       &\approx& {Z E \over 6\pi\sqrt{\phi_c}(1-\nu^2)} (\phi-\phi_c)^{3/2}. \label{Ywalton2}
\end{eqnarray}
For ``infinitely rough [{\it sic}]'' spheres that support any amount of shear traction across the contact area, the elastic stress is larger by a factor of $(3-2\nu)/(2-\nu)$.  Walton also finds the anisotropy of the stress: $\sigma_{xx}/\es=1/4$ for smooth spheres and $\nu/(12-8\nu)$ for rough spheres.

The connection between particle and packing behavior was broadly explored by O'Hern et al.\ \cite{OHernPRE03} using numerical simulation.  There the particles are frictionless disks or spheres that repel with central force $F=(\varepsilon/d)(h/d)^a$, where $\varepsilon$ is a microscopic energy scale, $d$ is the particle diameter, and $h$ is the compression.  Different $a$, different dimensionality, and different particle size distributions are all examined.   The pressure $P$, shear modulus $G$, bulk modulus $K$, and coordination number $Z$ are always found to scale with volume fraction as
\begin{eqnarray}
    P &=& P_o(\phi-\phi_c)^a            \label{Pepitome}\\
    G &=& G_o(\phi-\phi_c)^{a-1/2}  \label{Gepitome}\\
    K &=& K_o (\phi-\phi_c)^{a-1}       \label{Kepitome}\\
    Z &=& Z_c + Z_o(\phi-\phi_c)^{1/2}  \label{Zepitome}
\end{eqnarray}
Here $K_o=a \phi P_o$ follows from $K \equiv -V\partial P/\partial V=\phi \partial P/\partial \phi$, $Z_c$ equals twice the dimensionality of the system, and $\phi_c=0.639\pm0.001$ for monodisperse three dimensional spheres.  Some of these scaling relations were found in previous \cite{BubbleModelPRL, BubbleModelPRE, MaksePRL99} and subsequent \cite{MaksePRE04, EllenbroekPRL06} simulations for special cases.  An important general conclusion from all these studies is that the microscopic deformation is nonaffine -- the particle positions adjust away from the macroscopic deformation field in order to reduce individual compression.  This causes the scaling exponent of the shear modulus to be $+1/2$ larger than the affine expectation, so that resistance to shear is dramatically lower near $\phi_c$.  By contrast the exponent for the bulk modulus is unchanged, but $K_o$ is typically 10-30\% smaller than if the motion were affine.  These findings suggest that Walton's prediction, Eq.~(\ref{Ywalton1}), is an overestimate.

It is instructive to directly compare the pressure prediction of Walton with the simulations of O'Hern, et al.  For monodisperse Hertzian elastic spheres in 3-dimensions, $a=3/2$, $\phi_c=0.64$, and $Z_c=6$,  the simulation results quoted in Table~I of \cite{OHernPRE03} are $P_o=0.35(\varepsilon/d^3)$,  $G_o=0.14(\varepsilon/d^3)$, and $Z_o=7.7$.  Agreement of the simulation force law with the Hertz Eq.~(\ref{Fhertz}) requires that the energy scale be taken as $\varepsilon=Ed^3/[3(1-\nu^2)]$.  Altogether this gives the expectation based on the O'Hern et al.\ simulation as $P=0.117[E/(1-\nu^2)](\phi-\phi_c)^{3/2}$.  As expected due to nonaffine motion, this is slightly smaller (by a factor of 0.76) than Walton prediction Eq.~(\ref{Pwalton2}) evaluated with $Z=6$ and $\phi_c=0.64$.  Therefore, if the particles adjust under uniaxial compression so that the stress becomes isotropic, a lower bound on $\es$ would be about 0.76 times Walton's pressure prediction; this corresponds to 0.29 times his Eq.~(\ref{Ywalton1}) prediction for $Y$.  In the absence of further guidance, we will simply use Eq.~(\ref{Ywalton1}) to deduce $E$ from measurements of $Y$; the actual Young modulus could be up to three times larger.

\section{Experiment}

\begin{figure}
\includegraphics[width=2.75in]{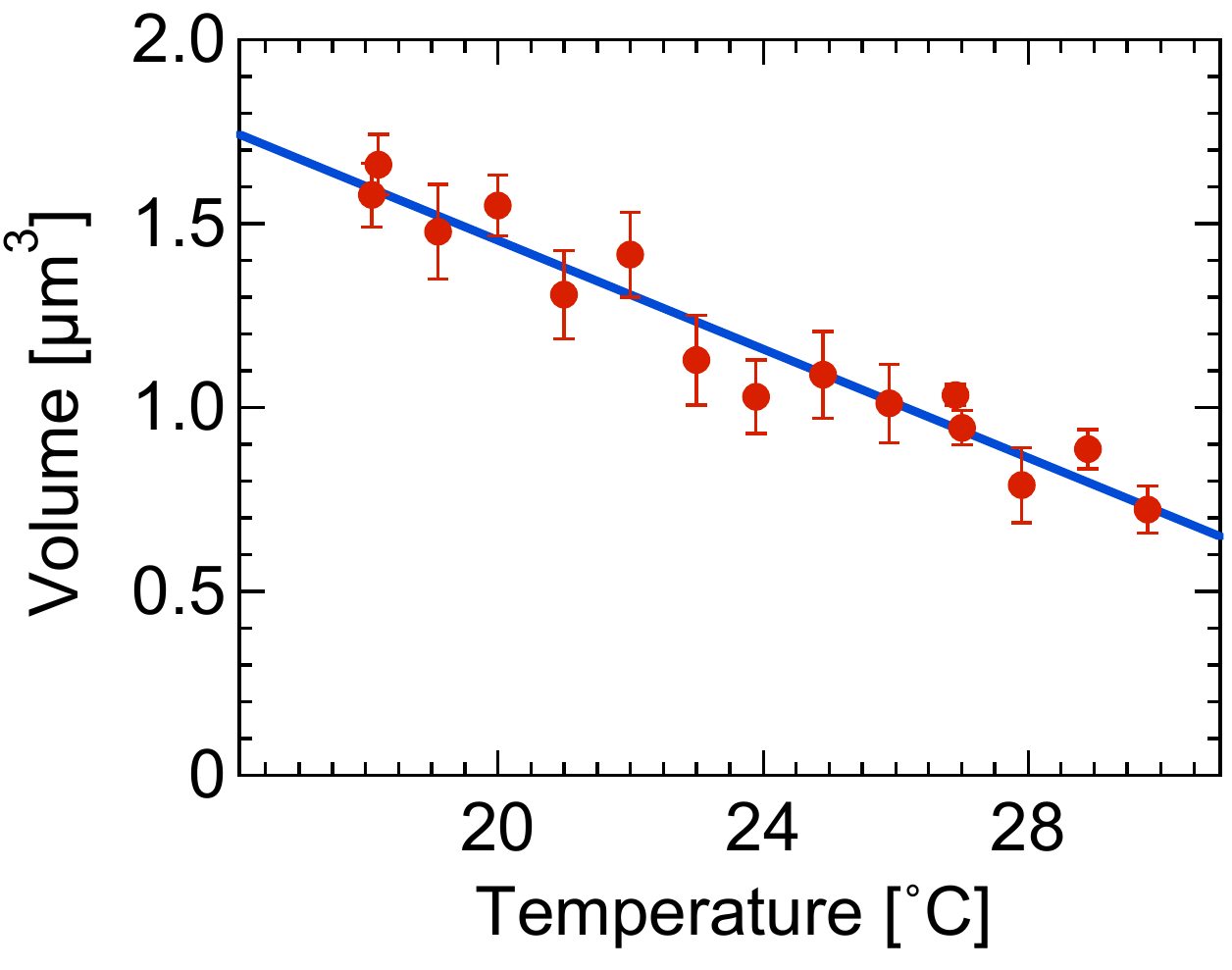}
\caption{(Color online) Particle volume $V$ versus temperature $T$, determined by dynamic light scattering from a dilute sample.  The line is an empirical fit, $V=(2.93~\mu{\rm m}^3)[1-T/(39.6^\circ{\rm C})]$.  Above $35^\circ{\rm C}$ the volume collapses to $0.092~\mu$m.  }
 \label{particlesize}
\end{figure}

\begin{figure}
\includegraphics[width=2.75in]{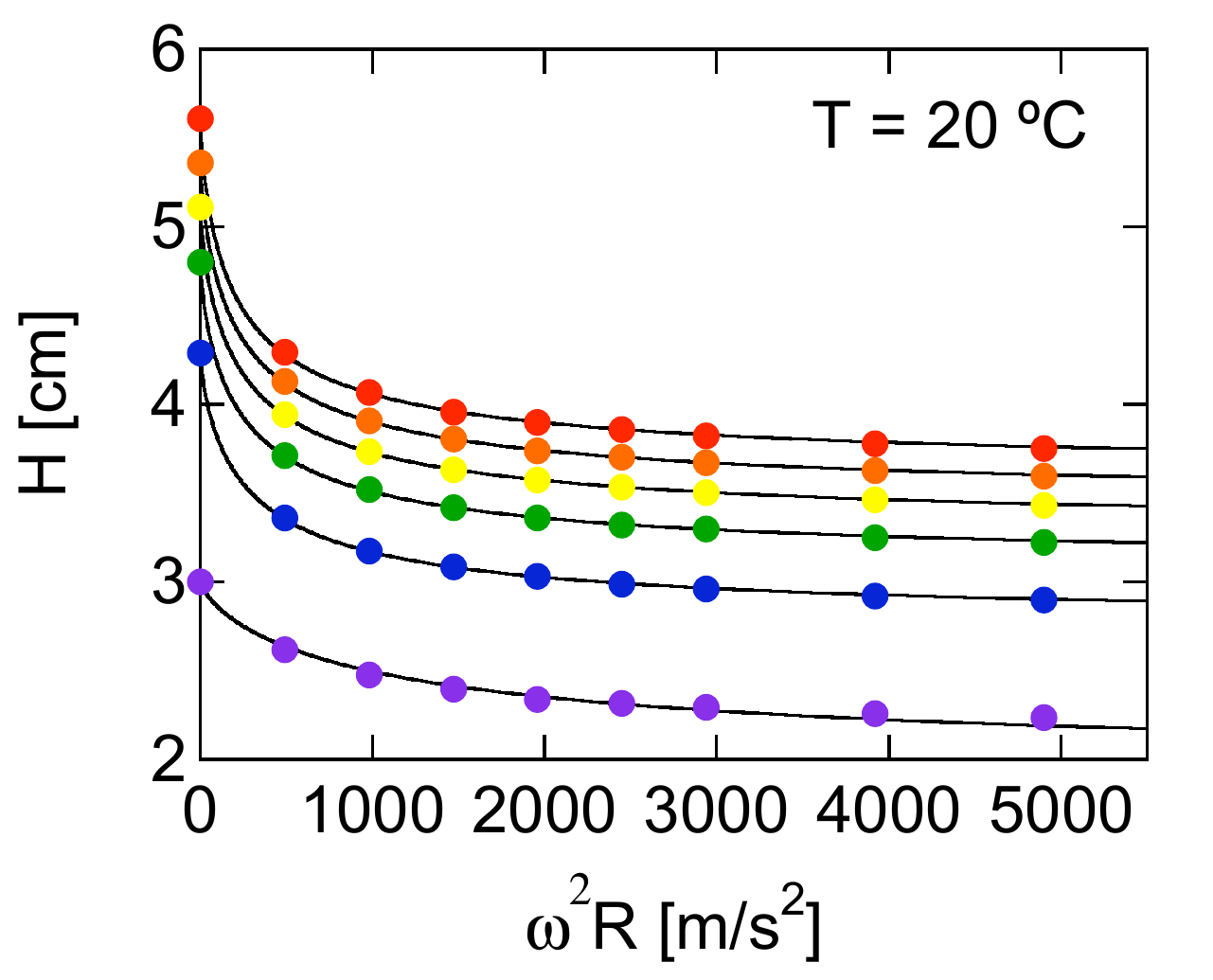}
\caption{(Color online) Radial packing height $H$ versus radial acceleration for samples at $T=20^\circ$C but with different initial heights, $H_c$, as indicated by the points along the y-axis.  The curves represent fits to Eq.~(\protect{\ref{Hhertzianmax}}) with $H_c$ held fixed.  The statistical uncertainty in measuring $H$ is about 0.1~mm, much less than the symbol size.} \label{20c}
\end{figure}

\begin{figure}
\includegraphics[width=2.75in]{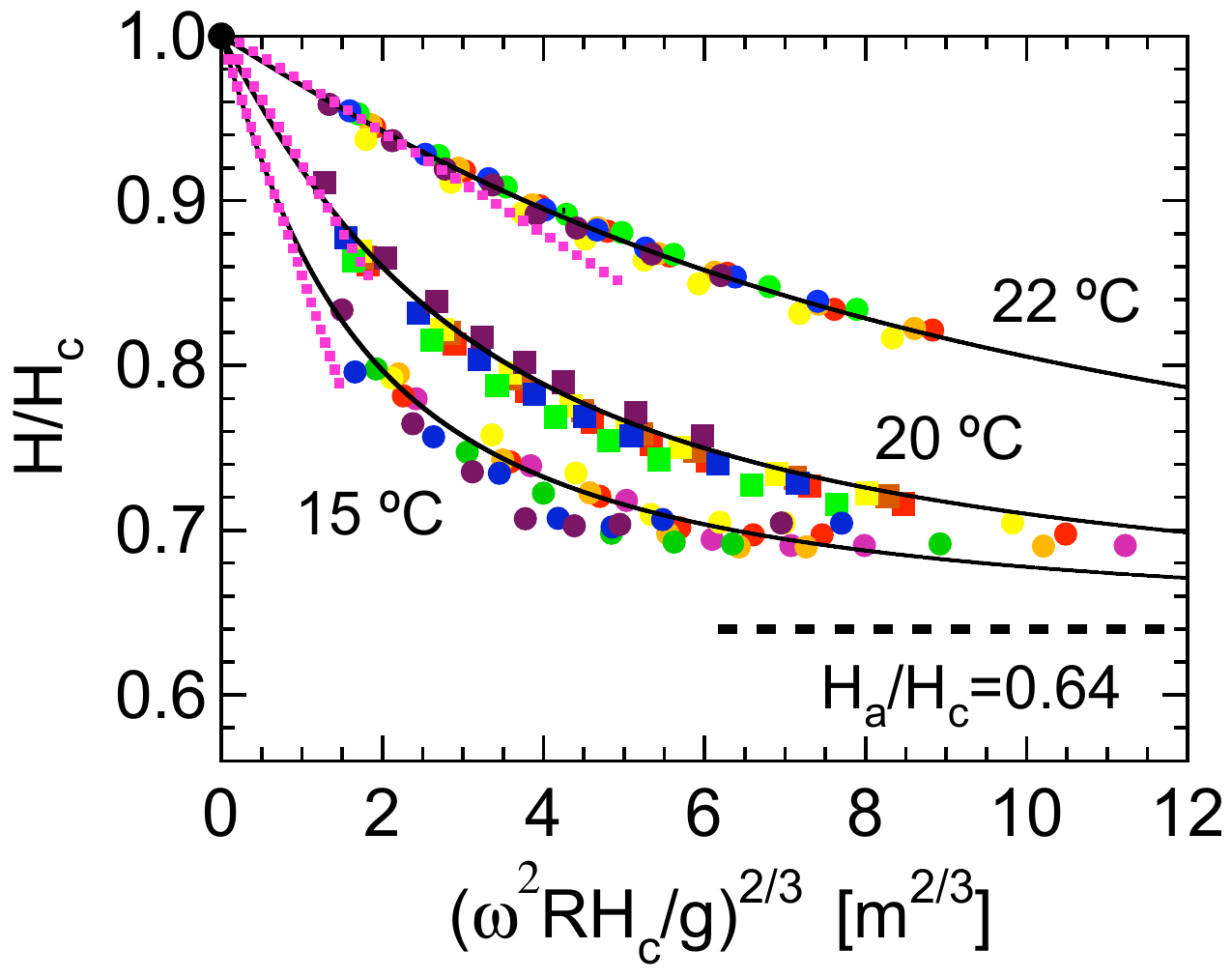}
\caption{(Color online) Packing height versus acceleration, with both axes scaled so as to cause collapse, for samples at different temperatures, as labeled.  Different color symbols represent different initial sample heights as in Fig.~\protect{\ref{20c}}.  The pink dotted lines are the initial decays, which are linear on such a plot for Hertzian particles.  The solid curves are fits to Eq.~(\protect{\ref{Hhertzianmax}}), and asymptote to $0.64\pm0.01$ as indicated by the horizontal dashed line.} \label{collapse}
\end{figure}

\begin{figure}
\includegraphics[width=2.75in]{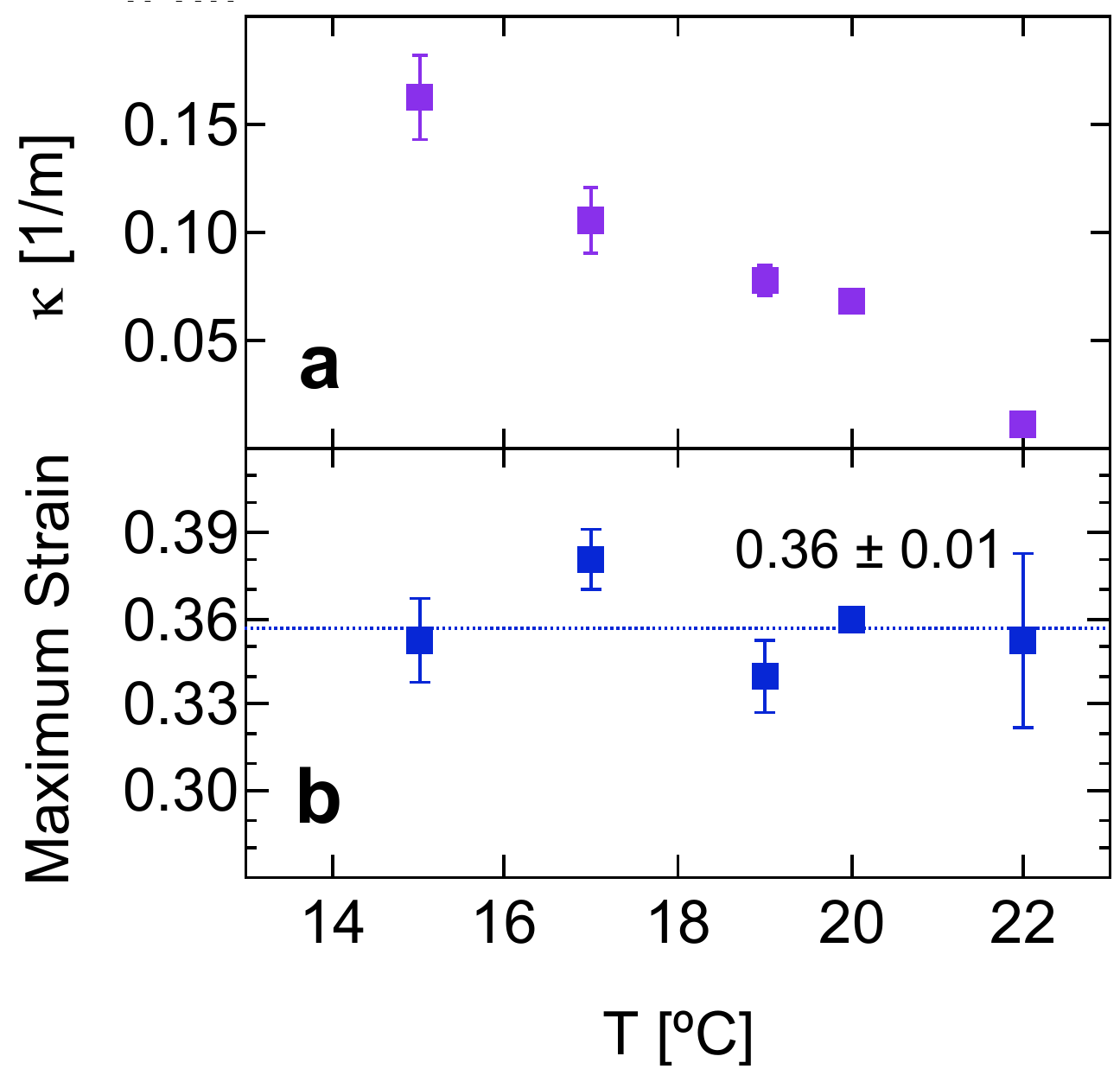}
\caption{(Color online) Temperature dependence of the parameters $\kappa$ and maximum strain $\gamma_m$, obtained from fits of Eq.~(\protect{\ref{Hhertzianmax}}) to normalized compression data as demonstrated in Fig.~\protect{\ref{collapse}}.  The assumed stress-strain relation, Eq.~(\protect{\ref{Phertzianmax}}), is Hertzian at low strains, $\es=Y\gamma^{3/2}$, and the sample resists compression beyond a maximum strain $\gamma_m$. The sample elasticity scales as $Y=\phi_c\Delta\rho g/\kappa$.  The $\gamma_m$ results are constant to within uncertainty, and average to $0.36\pm0.01$ as indicated by the horizontal dashed line.   The error bars are set by the degree of collapse and accuracy of the fits in Fig.~\protect{\ref{collapse}}.} \label{fitparams}
\end{figure}

\begin{figure}
\includegraphics[width=2.75in]{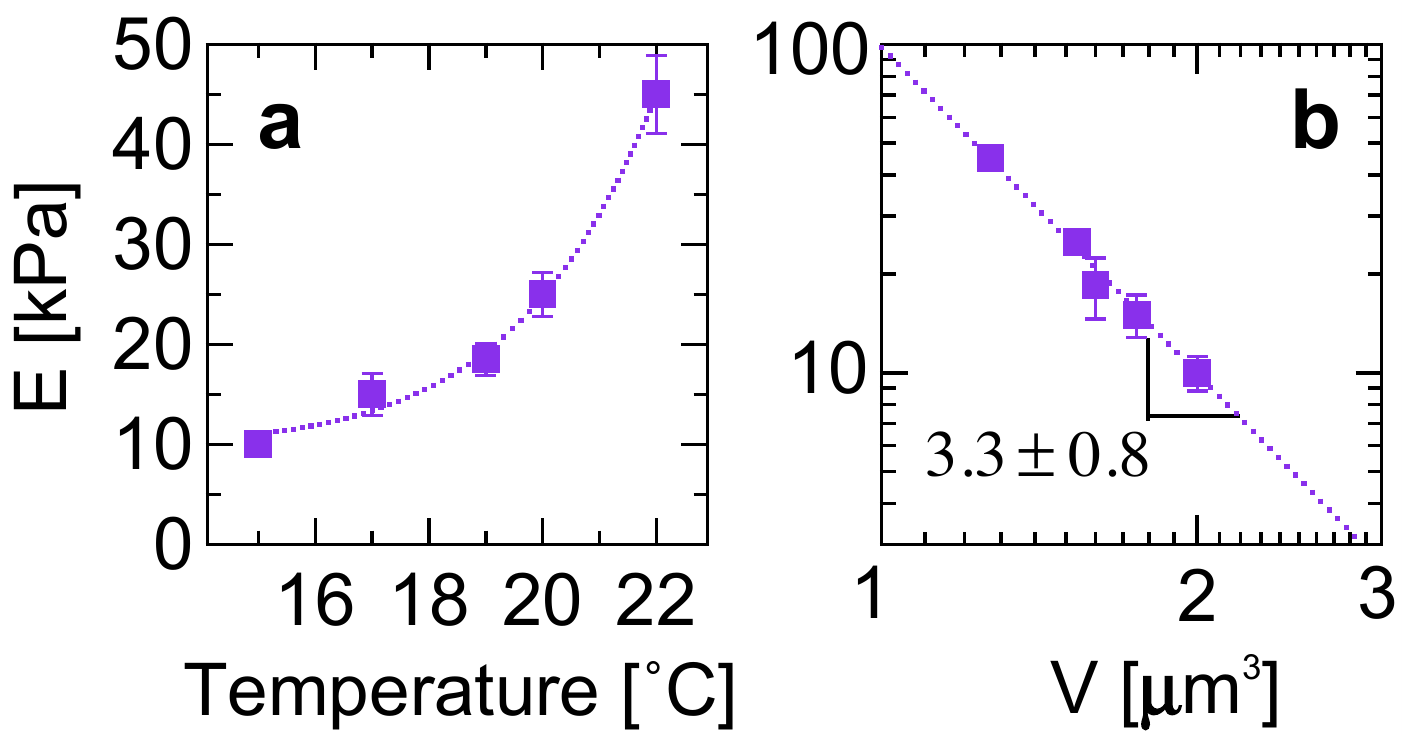}
\caption{(Color online) Young's elastic modulus $E$ of the particulate material plotted versus (a) temperature and (b) particle volume.  The values of $E$ are deduced from the fitting parameters $\kappa$ in Fig.~\protect{\ref{fitparams}}a using $Y=\phi_c \Delta\rho g/\kappa$ and Walton's relation in Eq.~(\ref{Ywalton1}).   The dashed line in (b) is a power-law, with slope as labeled.  It translates to the dashed curve in (a) using the empirical fit to particle volume versus temperature in Fig.~\ref{particlesize}. } \label{EvsTV}
\end{figure}

In this section we use the above theory to analyze the elasticity of a suspension of N-isopropylacrylamide (NIPA) microgel beads.  The particles are synthesized via free-radical polymerization \cite{PeltonACIS00, alsayed05, reufer09, YodhNature09, YodhPRL10, yunker10}. Briefly, NIPA monomer (Acros) and methylene-bis(acrylamide) crosslinker (Polyscience, Inc.) are mixed in aqueous solution. Ammonium persulfate (Fisher) is then injected into solution to initiate the polymerization. Spherical particles are formed within an hour. Next these are thoroughly washed and redistributed in a 1~mM sodium dodecylsulfate solution.  In Ref.~\cite{Nordstrom2010} we reported on microfluidic measurements of the shear rheology for dense suspensions of the very same samples.

For our samples the particle diameter is of order 1 micron in diameter.  But more importantly, as shown by the dynamic light scattering (DLS) results plotted in Fig.~\ref{particlesize}, the volume can decrease by more than a factor of 2 with a modest temperature increase.   Over the temperature range studied here, the particle volume exhibits an approximately linear dependence with temperature as noted in the figure caption.  At high temperatures, above approximately $35^\circ$C, the particles suddenly collapse to a temperature-independent volume of $0.092~\mu$m. For hard colloidal spheres sterically stabilized by graphted polymer, the ``hydrodynamic radius'' given by DLS is somewhat larger than the hard-sphere radius given by electron microscopy.  Here, the NIPA particles are sterically stabilized by dangling chains that emerge from crosslinking sites within the gel.  It is not possible to remove the beads from solution and measure their size by electron microscopy.  However, since the rheological features reported in Ref.~\cite{Nordstrom2010} exhibit power-law behavior in $\phi-\phi_c$ where $\phi_c=0.635\pm0.003$ coincides well with the volume fraction of randomly close-packed spheres, we conclude that the hydrodynamic radius given by DLS corresponds closely with the actual physical radius of the particles.

As the particles shrink with increasing temperature, the density of crosslinking sites necessarily increases and the elastic modulus hence increases in some way that we wish to determine.  Our experimental protocol is as follows.  We load different volumes of a stock suspension of particles into six 50~$\mu$L glass capillary tubes (ID = 0.8~mm, length = 10~cm), and seal at both ends with optical glue.   The number density of particles is $0.284/\mu{\rm m}^3$, as determined by counting particles in a three-dimensional confocal microscopy image of a sample with known dilution.  Loading is done at room temperature, where the particles occupy a volume fraction of approximately 40\%.  The sample volumes are chosen so that once the particles sediment to close packing, the range of initial {\it radial} heights $H_c$ (see Fig.~\ref{setup}) spans between about 3 and 6~cm.   This is long enough that the shape of the bottom of the tubes plays no role.  We load all the tubes into a thermostated centrifuge (Marathon 21000R), and let the particles settle at a fixed angular rotation speed $\omega$. This centrifuge has a rotation radius of $R=10$~cm and angle away from vertical of 30$^{\circ}$; the temperature range is about 10-25$^\circ$C and is held constant to $\pm0.1^{\circ}{\rm C}$.  To determine the time needed for complete settling to mechanical equilibrium, we measure the height versus time for a range of rotation rates by periodically removing the tubes and tracking the sediment-supernatant interfaces.  The settling equilibrates within a few hours for high rotation rates, but takes up to several days for low rotation rates. Once the samples have settled to their equilibrium heights we place the tubes in a holder on an optical table, and photograph using a digital Nikon D70 camera.  We analyze the images for the length of the sediment and convert to the radial height, $H$, via the appropriate trigonometric factor (See Fig.~\ref{setup}).  The statistical uncertainty is about $\Delta H=0.1$~mm.  Then we agitate the samples to redistribute the particles, and repeat at a different rotation rate.  It should be noted that the packing eventually rebounds elastically to random close packing, but this process occurs on a scale of many hours to days while our height measurements take only minutes.  We also note that the particles clearly return to their original spherical shape with no plastic deformation: we can repeat the experiment with a redistributed sample or a fresh sample and produce the same result.

Example data for radial height versus angular acceleration are shown in Fig.~\ref{20c}, for all six tubes at temperature $T=20^{\circ}{\rm C}$. The points at $\omega=0$ are not from centrifuge measurements, but instead are the expectations for $H_c$ based on particle number density, particle size, sample volumes, and a random close packing volume fraction of $\phi_c=0.64$.  Since the colloidal gel particles are presumably Hertzian, and since the height data do not decrease toward zero, we fit to Eq.~(\ref{Hhertzianmax}) keeping $H_c$ fixed.   These fits are all good, and interpolate smoothly between the expected $H_c$ at $\omega=0$ and the centrifuge results at $\omega>0$.  This gives confidence in our sample characterization and centrifugation measurement procedures.

We now attempt to collapse the compression data according to the general expectation in the theory section for any constitutive law.  In particular, in Fig.~\ref{collapse} we plot the normalized radial height $H/H_c$ versus the length $x=\omega^2 R H_c/g$ raised to the 2/3 power, for all six tubes and for three different temperatures.  As such, the y-axis represents the observation while the x-axis consists of a combination of the three control parameters that could be varied in experiment.   We see in Fig.~\ref{collapse} that this does indeed cause good collapse of the height data at each temperature.  Also as expected, at higher temperatures the particles are smaller and hence stiffer and less compressed.  Furthermore the initial decay is linear on such a plot, in accord with Hertzian behavior at small strains.

Next we fit the collapsed data, for all tubes at a given temperature, to the prediction Eq.~(\ref{Hhertzianmax}) for a sphere packing that is Hertzian at small strains and that cannot be strained beyond some maximum $\gamma_m$.  The first fitting parameter is the reciprocal length, $\kappa$, defined as discussed earlier so that $H_c k = \kappa x$ in Eq.~(\ref{Hhertzianmax}); the value of $\kappa$ is determined by the elasticity of the medium and sets the slope of the initial linear decay seen in Fig.~\ref{collapse}.  The second fitting parameter is the maximum strain $\gamma_m$, which sets the asymptotic value of the scaled height as $H_a/H_c = 1-\gamma_m$ at high rotation speed.  The fits at all temperature are as satisfactory as those shown in Fig.~\ref{collapse}.  Therefore the elasticity of the packing is adequately described by the empirical form of Eq.~(\ref{Phertzianmax}), and the fitting parameters have the intended physical meaning.

The fitting parameters $\kappa$ and $\gamma_m$ are plotted versus temperature in Fig.~\ref{fitparams}.  The top plot shows how $\kappa$ decreases with temperature, and is analyzed in the following paragraph.  The bottom plot shows that the maximum strain is independent of temperature to within experimental uncertainty, and averages to $\gamma_m=0.36\pm0.01$.  This gives an asymptotic relative height of $H_a/H_c=1-\gamma_m=0.64$ as shown by the dashed line in Fig.~\ref{collapse}.   Taking the random close packing fraction as $\phi_c=0.64$ and assuming the particles deform at fixed volume without deswelling, the volume fraction $\phi_m$ at maximum strain would be given by $\phi_cH_c=\phi_mH_a$ as $\phi_m=1$.   This is space filling, and implies than any compression $H/H_c$ below the dashed line at $1-\gamma_m=0.64$ in Fig.~\ref{collapse} could only be accomplished by deswelling.  Since the actual data at all temperatures are always above this limit, and appear to approach it smoothly from above, we conclude that the particles do not deswell and are effectively incompressible.  This is consistent with reports that the Poisson ratio of bulk samples of swollen NIPA is close to 1/2 \cite{Hirotsu1991, LinJCP2007}.  It is also consistent with the observation that $\approx 10$~MPa of applied pressure is needed for noticeable deswelling of NIPA microgels \cite{LietorSantos09}, while the pressure here does not exceed 0.03~MPa.  Stated differently, we estimate that a radial acceleration of about $10^6$~m/s$^2$ (200 times our maximum) would be required to induce deswelling and, thus, to cut off the divergence assumed in Eq.~(\ref{Phertzianmax}).  See Ref.~\cite{KnaebelPGN1997b, VervoortPolymer05} for the deswelling of polyelectrolyte gels under compression.

We now analyze the $\kappa$ results in Fig.~\ref{fitparams}a for the elasticity of the NIPA particles.  Recall that the Hertzian elastic constant in $\es=Y\gamma^{3/2}$ is given by Eq.~(\ref{Ykappa}) as $Y=\phi_c \Delta\rho g/\kappa$.  Here the density difference $\Delta\rho$ between particles and water at $T=20^\circ$C is found to be 0.08~g/cm$^3$ by measuring the terminal sedimentation speed of single spheres at room temperature, and equating gravity to Stokes forces.  Density values for different temperatures are then deduced using the particle volume versus temperature data of Fig.~\ref{particlesize}.  After converting $\kappa$ to $Y$,  we finally deduce the Young elastic modulus according to Eq.~(\ref{Ywalton1}) as $E=6\pi(1-\nu^2)Y/(\phi Z)$, using $\nu=1/2$ and evaluating the denominator at random-close packing, $\phi=\phi_c=0.64$ and $Z=6$.  The results are plotted in Fig.~\ref{EvsTV} versus (a) temperature and (b) particle volume.  As expected, $E$ increases with temperature since the particles shrink.  The order of magnitude is tens of kPa, as found previously for bulk NIPA gel samples \cite{Hirotsu1991, TakigawaPGN1997, MatzelleJPCB2002, MatzelleMacro2003, LinJCP2007}.

The scaling of $E$ vs $V$ plotted in Fig.~\ref{EvsTV}b is fit well by a power-law of $E\sim 1/V^{3.3\pm0.8}$.  Since the number of cross links in the gel beads does not change as they shrink with temperature, the same exponent $x={3.3\pm0.8}$ holds for the scaling of $E$ with cross-link density.  For this Flory theory applied to neutral gel \cite{deGennes, RubinsteinColby, ObukhovRubinstein} predicts a value of $x=2.25$ in good solvent and $x=3$ in poor (theta) solvent. Experimental values mentioned in \cite{ObukhovRubinstein} are $x = 2.3$ and $2.4$ for good solvents and $x = 3.0$ and 3.7 for theta solvents. In our case, as the temperature is increased, not only cross-link density increases but also, solvent quality decreases \cite{Hirotsu1991}, which leads to an increased exponent value. Furthermore, Flory theory supposes that the gel network deswells in an affine way. Affine deformation can be impaired two-fold: heterogeneities in the cross-links distribution may arise from the synthesis mechanisms \cite{CandauPGN1996, ObukhovRubinstein}, while entanglements between cross-links may be created as the network shrinks. Departure from affine behavior results in an increase in the $E$ versus cross-link density exponent beyond the $x=2.25$ value \cite{ObukhovRubinstein}. Experimental evidence for such a non-affine shrinking of NIPA gels were reported in \cite{LinJCP2007} where a harder skin is found to form at their surface upon temperature increase.  In our case, such a skin would have to be thicker than the scale of deformation under compressions in the Hertzian regime.  And finally, our samples are not perfectly neutral as assumed for the Flory value of $x=2.25$; rather, the NIPA beads have a very slight negative charge, and the surfactant is anionic.  This situation can be compared with polyelectrolyte gels swollen in brine, where Hertzian behavior and $x=3.5$ was reported in Ref.~\cite{KnaebelPGN1997} from compression of individual beads.  Altogether, the decrease of solvent quality and the possibilities of non-affine shrinkage and of charging effects are all consistent with our experimental measure of $x=3.3\pm0.8$ for the scaling exponent of $E$ with cross-link density.

\section{Conclusion}

Here we presented both theory and experiment for the compression of an elastic medium under centrifugation.  The formalism culminates in Eq.~(\ref{soln}) for the total sample height expressed in terms of a function $f(s)$ defined by Eqs.~(\ref{parts2}-\ref{parts3}).  Remarkably, this represents an exact solution for arbitrary stress-strain elastic constitutive law and was found without having to first solve explicitly for the height dependence of the compressive strain, which is greatest at the bottom of the sample and decreases to zero at the top.  We hope that the example solutions developed for power-law media, and for harmonic and Hertzian media with a maximum strain, will be of use to experimentalists.  These same predictions hold for media that are so soft that gravity causes measurable compression, by replacing $\omega^2 R$ with $g$.

To illustrate, and to characterize the elastic nature of particles in a suspension of thermoresponsive gel beads of interest for shear rheology experiments \cite{Nordstrom2010}, we presented a series of measurements of sample height versus rotation speed, filling height, and temperature.  The results are in good agreement with Eq.~(\ref{Hhertzianmax}), the exact solution for a Hertzian medium with maximum strain as specified by the constitutive law of Eq.~(\ref{Phertzianmax}).  The data thus are analyzed in terms of the linear elastic Young modulus $E$ of the gel material, and in terms of the maximum strain $\gamma_m$ beyond which the medium cannot be compressed.  The former is demonstrated to vary as a large power of cross-link density, and was used in Ref.~\cite{Nordstrom2010} to non-dimensionalize shear rheology data.  The latter is a constant, $\gamma_m=0.36\pm0.01$, whose value implies that the gel beads are incompressible and do not deswell when deformed; this conclusion is important in Ref.~\cite{Nordstrom2010} for knowing the packing fraction above jamming.  Valuable particle-scale information such as this is now straightforward to obtain by centrifugal compression measurements, analyzed using the theory presented here.

\begin{acknowledgments}
We thank P. E. Arratia, F. Lequeux, A.J. Liu, and A. G. Yodh for helpful conversations; we thank A. Alsayed, A. Basu, and Z. Zhang for helping synthesize the particles.  This work was supported by the National Science Foundation through grants MRSEC/DMR05-20020 and DMR-0704147.
\end{acknowledgments}


\bibliography{CentComp_References}


%
\end{document}